\documentclass{article}

\PassOptionsToPackage{square, numbers}{natbib}

\usepackage[final]{neurips_2019}

\usepackage[utf8]{inputenc} 
\usepackage[T1]{fontenc}    
\usepackage{hyperref}       
\usepackage{url}            
\usepackage{booktabs}       
\usepackage{amsfonts}       
\usepackage{nicefrac}       
\usepackage{microtype}      
\usepackage{graphicx}

\title{Evolution-based Fine-tuning of CNNs for Prostate Cancer Detection}

\author{%
  Khashayar Namdar \\
  Lunenfeld-Tanenbaum Research Institute\\
  Sinai Health System\\
  \texttt{knamdar@lunenfeld.ca} \\
  \And
  Isha Gujrathi\\
  Brigham and Women's Hospital\\
  Boston, United States\\
  \And
   Masoom A. Haider\\
  University of Toronto \\
  Lunenfeld-Tanenbaum Research Institute\\
  Sinai Health System\\
  \And
   Farzad Khalvati \\
  University of Toronto \\
  Lunenfeld-Tanenbaum Research Institute\\
  Sinai Health System\\
   \texttt{farzad.khalvati@utoronto.ca} \\
}

\begin{document}

\maketitle

\begin{abstract}
 Convolutional Neural Networks (CNNs) have been used for automated detection of prostate cancer where Area Under Receiver Operating Characteristic (ROC) curve (AUC) is usually used as the performance metric. Given that AUC is not differentiable, common practice is to train the CNN using a loss functions based on other performance metrics such as cross entropy and monitoring AUC to select the best model. In this work, we propose to fine-tune a trained CNN for prostate cancer detection using a Genetic Algorithm to achieve a higher AUC. Our dataset contained 6-channel Diffusion-Weighted MRI slices of prostate. On a cohort of 2,955 training, 1,417 validation, and 1,334 test slices, we reached test AUC of 0.773; a 9.3\% improvement compared to the base CNN model.
\end{abstract}

\section{Introduction}

Prostate cancer detection is a longstanding challenge in medical imaging for which, many algorithms have been proposed using conventional machine learning algorithms~\cite{Khalvati2015a,Khalvati2016,Khalvati2018,Hussain2018}. With the advent of deep Convolutional Neural Networks (CNNs) and the promising results achieved by CNNs in computer vision tasks, there has been a shift in designing computer-aided detection algorithms for prostate cancer toward CNN architectures~\cite{Kwak2017,Song2018,Wang2017,Yang2017,Yoo2019}. From Machine learning perspective, prostate cancer detection is a binary classification task. To evaluate performance of such a binary classification model, Area Under Receiver Operating Characteristic (ROC) curve (AUC) is usually used. In medical imaging in particular, AUC is widely used as a performance measure~\cite{Park2004}.

Conventional approach for training a CNN is backpropagation~\cite{nref1}. For a loss function to work in backpropagation, it must be differentiable~\cite{nref2}. However, AUC is not differentiable and therefore, CNNs are usually trained using a loss functions based on other performance metrics such as cross entropy. During the training process, while loss is being minimized, AUC is monitored and the best performing model is selected based on the highest AUC~\cite{Wang2017,Yang2017,Yoo2019}. The challenge here is that a model optimized for minimum loss may not necessarily produce the best possible AUC. To address this issue, we propose an evolution-based method to fine-tune a CNN that has been trained for prostate cancer detection. 

Genetic algorithms (GAs) are a class of evolutionary methods which have been used for optimization in machine learning for a number of years~\cite{ref9, DBLP:journals/corr/XieY17}. GAs do not rely on the derivative of the loss function (called fitness function in evolutionary algorithms domain). High computational cost of GAs has limited their application in CNN optimization~\cite{Castillo_g-prop-iii:global}. Nevertheless, there are efforts for using GAs for optimizing CNNs for image classification~\cite{ref10,Sun2018}. 

In this paper, we use a GA to fine-tune a CNN, which has been trained for prostate cancer detection using Diffusion-Weighted Magnetic Resonance Imaging (DW-MRI). The GA is applied to the fully connected (FC) layers of the CNN, thus the computational cost is significantly reduced. Although more sophisticated CNN architectures have been used for prostate cancer detection~\cite{Wang2017,Yang2017,Yoo2019}, we developed a simple CNN with 3 convolutional layers and 3 FC layers to demonstrate capability of the proposed method in improving the performance of CNN architectures. The proposed evolutionary fine-tuning algorithm improves AUC of the CNN by 9.3\% in the test set, which includes 1,334 slices of DW-MRI images of prostate.

\section{Methods}

DW-MRI images from 414 prostate cancer patients (5,706 2D slices) were used as the dataset. Institutional review board approval was obtained for this study. Six DWI sequences were included for each slice: apparent diffusion coefficient (ADC) map, and five different b-value images (0, 100, 400, 1000, and 1600 $sm^{-2}$). Images were preprocessed and prostate regions were cropped using manual contours of the prostate. Each prostate region was resized to $64 \times 64$ pixels. The dataset was divided into training (217 patients, 2,955 slices), validation (102 patients, 1,417 slices), and test sets (95 patients, 1,334 slices). Label for each slice was generated based on the targeted biopsy results where a clinically significant prostate cancer (Gleason score>6) was considered a positive label.

\begin{figure}[htp]
  \centering
  \includegraphics[scale=0.45]{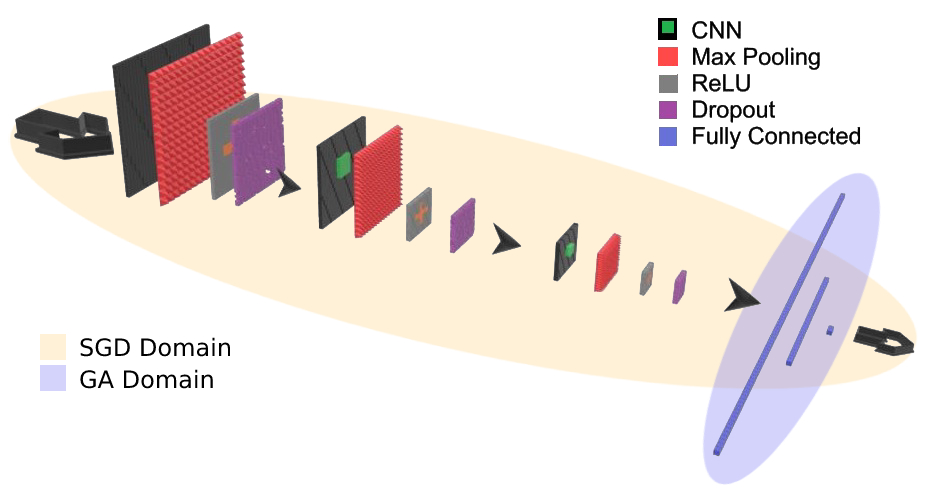}
  \caption{The Proposed GA-CNN Architecture}
  \label{figure:1}
\end{figure}

Figure~\ref{figure:1} shows the CNN architecture that we used for the experiments. The configuration of the CNN is shown in Table~\ref{table:1}. Padding was not used in the architecture and stride was equal to one. Weights of CNN layers were initialized by Xavier method~\cite{Glorot10understandingthe}. All biases and weights of FC layers were randomly initialized from a uniform distribution over [0, 1]. The model was trained based on Cross Entropy loss function and optimized by Stochastic Gradient Descent (SGD)~\cite{Ruder2016}. We used Python 3.7.3, PyTorch 1.1.0, and Ubuntu 19.04 for the experiments.

\begin{table}
  \caption{Configuration of the CNN}
  \label{table:1}
  \centering
  \begin{tabular}{ll}
    \toprule
    Layer     & Configuration \\
    \midrule
    CNN-1	&input channels=6, output channels=16, kernel size=7 \\
    Max Pooling-1	&kernel size=2 \\
    Dropout-1	&probability=0.1\\
    CNN-2	&input channels=16, output channels=32, kernel size=5\\
    Max Pooling-2	&kernel size=2\\
    Dropout-2	&probability=0.1\\
    CNN-3	&input channels=32, output channels=64, kernel size=4\\
    Max Pooling-3	&kernel size=2\\
    Dropout-3	&probability=0.1\\
    Fully Connected-1	&input size=1024, output size=256\\
    Fully Connected-2	&input size=256, output size=64\\
    Fully Connected-3	&input size=64, output size=2\\
    \bottomrule
  \end{tabular}
\end{table}

Convolution layers can be considered as feature extractors, which are optimized by SGD. However, FC layers are in fact classifiers, which may not reach an optimum point in terms of AUC by SGD. Thus, we hypothesize that by introducing a GA to FC layers, the classifier portion of the CNN is further optimized for AUC. Therefore, our proposed approach is similar to freezing shallow layers (feature extractors) of CNNs in Transfer Learning~\cite{Shie2015}.

The initial population of our GA includes 512 instances of the classifier (3 FC layers of the CNN). One instance is extracted from the trained CNN model while the remaining 511 instances are randomly initialized. Classifiers (instances) are ranked based on their AUC performance for the training set. Top half of the instances are transferred to the next generation. They are then crossed over and mutated to produce two remaining quarters of the generation. Mutation occurs with probability of 1\%. As long as the maximum AUCs on the validation and training sets are improved, this process continues.

\begin{figure}[htp]
  \centering
  \includegraphics[scale=0.25]{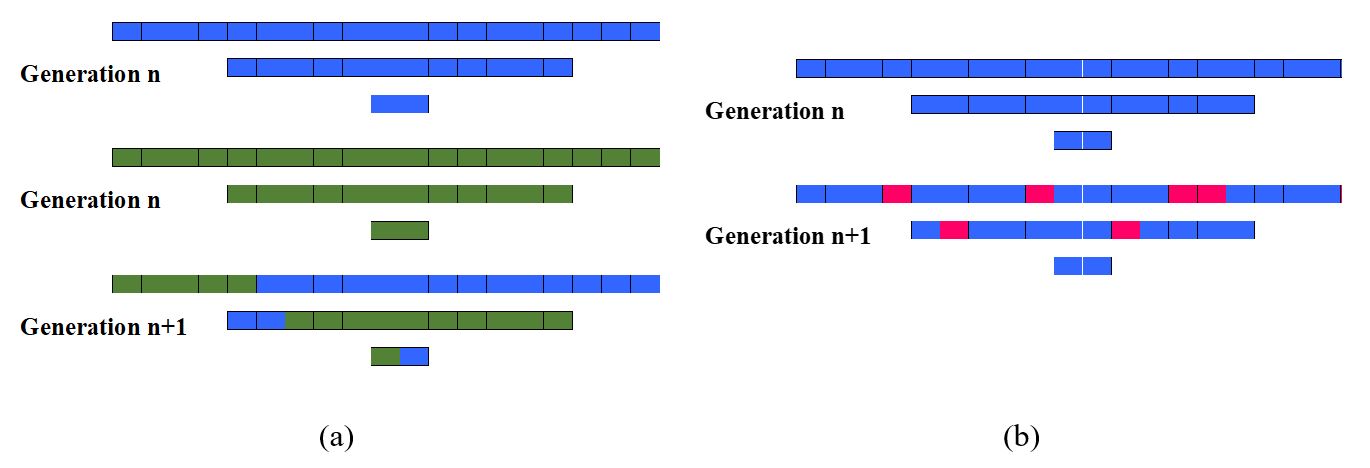}
  \caption{High-level GA application: (a) Crossover (b) Mutation}
  \label{figure:2}
\end{figure}

Even with targeting FC layers, computational cost and memory requirements of the GAs are still high. To mitigate this, we applied the crossover and mutation operations at layer level, instead of individual nodes (Figure~\ref{figure:2}). In other words, we do not optimize each individual weight and bias of the classifier and instead, parents and offsprings are in the form of an entire layer.

\section{Evaluation and Results}
Based on a grid search performed for selecting optimum hyper-parameters, we set learning rate to 0.001, and momentum equal to 0.8. L2 penalty of 0.001 and batch size of 1 were used. Although the maximum epoch number was 50, the best AUC was achieved in 10$^{th}$ epoch. Once he CNN was optimized by SGD, the model attained an AUC of 0.794 on the validation set and 0.707 on the test set. Our best model after applying the GA was achieved in the third generation. The AUC performance was 0.815 and 0.773 on the validation and test sets, respectively, which is a 9.3\% of AUC improvement on the test set. Table~\ref{table:2} lists the results in detail. We ran the GA algorithm on GeForce GTX 1060 and it took 10 min to reach the optimal AUC result.


\begin{table}[h!]
  \caption{AUC performance for SGD and proposed GA-based Method}
  \label{table:2}
  \centering
  \begin{tabular}{lll}
    \toprule
	& SGD	&GA\\
    \midrule
    AUC on train set	&0.867	&0.877\\
    AUC on validation set	&0.794	&0.815\\
    AUC on test set	&0.707	&0.773\\
    \bottomrule
  \end{tabular}
\end{table}


\section{Conclusion}
In this work, we proposed a GA-based method to fine-tune CNNs for prostate cancer detection. Monitoring validation set AUC during conventional training of CNNs results in a sub-optimal model. By applying a GA to FC layers, and performing crossover and mutation on the entire layer instead of individual coefficients, the proposed evolution-based fine-tuning procedure becomes feasible even for low-end GPUs such as GeForce GTX 1060. We demonstrated that for a simple CNN architecture with 3 convolutional layers and 3 FC layers, our proposed evolutionary algorithm can improve the AUC of test dataset by 9.3\%.

\bibliographystyle{unsrt}
\bibliography{refs}

\begin{thebibliography}{10}

\bibitem{Khalvati2015a}
F.~Khalvati, A.~Wong, and M.~A. Haider.
\newblock {Automated Prostate Cancer Detection via Comprehensive
  Multi-parametric Magnetic Resonance Imaging Texture Feature Models.}
\newblock {\em BMC medical imaging}, 15(1):27, 2015.

\bibitem{Khalvati2016}
F.~Khalvati, J.~Zhang, A.~Wong, and M.~A. Haider.
\newblock {Bag of Bags : Nested Multi Instance Classification for Prostate
  Cancer Detection}.
\newblock In {\em IEEE International Conference on Machine Learning and
  Applications (IEEE ICMLA)}, pages 146--151, 2016.

\bibitem{Khalvati2018}
F.~Khalvati, J.~Zhang, A.G. Chung, M.J. Shafiee, A.~Wong, and M.~A. Haider.
\newblock {MPCaD: A multi-scale radiomics-driven framework for automated
  prostate cancer localization and detection}.
\newblock {\em BMC Medical Imaging}, 18(1), 2018.

\bibitem{Hussain2018}
L.~Hussain, A.~Ahmed, S.~Saeed, S.~Rathore, I.~A. Awan, S.~A. Shah, A.~Majid,
  A.~Idris, and A.A. Awan.
\newblock {Prostate cancer detection using machine learning techniques by
  employing combination of features extracting strategies}.
\newblock {\em Cancer biomarkers : section A of Disease markers}, 2018.

\bibitem{Kwak2017}
J.~T. Kwak and S.~M. Hewitt.
\newblock {Nuclear Architecture Analysis of Prostate Cancer via Convolutional
  Neural Networks}.
\newblock {\em IEEE Access}, 2017.

\bibitem{Song2018}
Y.~Song, Y.~D. Zhang, X.~Yan, H.~Liu, M.~Zhou, B.~Hu, and G.~Yang.
\newblock {Computer-aided diagnosis of prostate cancer using a deep
  convolutional neural network from multiparametric MRI}.
\newblock {\em Journal of Magnetic Resonance Imaging}, 2018.

\bibitem{Wang2017}
X.~Wang, W.~Yang, J.~Weinreb, J.~Han, Q.~Li, X.~Kong, Y.~Yan, Z.~Ke, B.~Luo,
  T.~Liu, and L.~Wang.
\newblock {Searching for prostate cancer by fully automated magnetic resonance
  imaging classification: Deep learning versus non-deep learning}.
\newblock {\em Scientific Reports}, 2017.

\bibitem{Yang2017}
X.~Yang, C.~Liu, Z.~Wang, J.~Yang, H.~L. Min, L.~Wang, and K.~T. Cheng.
\newblock {Co-trained convolutional neural networks for automated detection of
  prostate cancer in multi-parametric MRI}.
\newblock {\em Medical Image Analysis}, 2017.

\bibitem{Yoo2019}
S.~Yoo, I.~Gujrathi, M.~A. Haider, and F.~Khalvati.
\newblock {Prostate Cancer Detection using Deep Convolutional Neural Networks}.
\newblock {\em arXiv}, abs/1905.1, 2019.

\bibitem{Park2004}
S.~Ho Park, J.~M. Goo, and C.~H. Jo.
\newblock {Receiver operating characteristic (ROC) curve: Practical review for
  radiologists}, 2004.

\bibitem{nref1}
H.~{Zhenlong}, Z.~{Qiang}, and W.~{Jun}.
\newblock The prediction model of air-jet texturing yarn intensity based on the
  cnn-bp neural network.
\newblock In {\em 2018 IEEE 3rd International Conference on Cloud Computing and
  Big Data Analysis (ICCCBDA)}, pages 116--119, April 2018.

\bibitem{nref2}
L.~{Zhao}, M.~{Mammadov}, and J.~{Yearwood}.
\newblock From convex to nonconvex: A loss function analysis for binary
  classification.
\newblock In {\em 2010 IEEE International Conference on Data Mining Workshops},
  pages 1281--1288, Dec 2010.

\bibitem{ref9}
D.~{Etter}, M.~{Hicks}, and K.~{Cho}.
\newblock Recursive adaptive filter design using an adaptive genetic algorithm.
\newblock In {\em ICASSP '82. IEEE International Conference on Acoustics,
  Speech, and Signal Processing}, volume~7, pages 635--638, May 1982.

\bibitem{DBLP:journals/corr/XieY17}
L.~Xie and A.~L. Yuille.
\newblock Genetic {CNN}.
\newblock {\em CoRR}, abs/1703.01513, 2017.

\bibitem{Castillo_g-prop-iii:global}
P.~A. Castillo, J.J. Merelo, J.~González, A.~Prieto, V.~Rivas, and G.~Romero.
\newblock G-prop-iii: Global optimization of multilayer perceptrons using an
  evolutionary algorithm.

\bibitem{ref10}
A.~{Rikhtegar}, M.~{Pooyan}, and M.~T. {Manzuri-Shalmani}.
\newblock Genetic algorithm-optimised structure of convolutional neural network
  for face recognition applications.
\newblock {\em IET Computer Vision}, 10(6):559--566, 2016.

\bibitem{Sun2018}
Y.~Sun, B.~Xue, M.~Zhang, and G.~G. Yen.
\newblock {Automatically Designing CNN Architectures Using Genetic Algorithm
  for Image Classification}.
\newblock {\em arXiv}, pages 1--12, 2018.

\bibitem{Glorot10understandingthe}
X.~Glorot and Y.~Bengio.
\newblock Understanding the difficulty of training deep feedforward neural
  networks.
\newblock In {\em In Proceedings of the International Conference on Artificial
  Intelligence and Statistics (AISTATS’10). Society for Artificial
  Intelligence and Statistics}, 2010.

\bibitem{Ruder2016}
S.~Ruder.
\newblock An overview of gradient descent optimization algorithms.
\newblock {\em arXiv}, 2016.

\bibitem{Shie2015}
C.~K. Shie, C.~H. Chuang, C.~N. Chou, M.~H. Wu, and E.~Y. Chang.
\newblock {Transfer representation learning for medical image analysis}.
\newblock In {\em Proceedings of the Annual International Conference of the
  IEEE Engineering in Medicine and Biology Society, EMBS}, 2015.

\end{thebibliography}

\end{document}